\documentclass[letterpaper, 10 pt, conference]{ieeeconf}  

\IEEEoverridecommandlockouts                              

\usepackage{graphicx}
\usepackage{amsmath}
\usepackage{amssymb}
\usepackage{comment}
\usepackage{color}
\usepackage{float}
\usepackage{mathrsfs}
\usepackage{hyperref}
\usepackage{cite}
\usepackage{enumitem}

\usepackage{amsfonts,latexsym,amsmath,amssymb,amsthm}
\usepackage[mathscr]{eucal}
\usepackage{graphicx,color}
\usepackage{comment}
\usepackage{hyperref}
\usepackage[margin = 1in]{geometry}

\makeatletter
\let\theoremstyle\@undefined                        
\makeatother

\usepackage{amsthm}

\newtheorem{nntheorem}{\bf Theorem}
\newtheorem{nnassumption}{\bf Assumption}
\newtheorem{nndefinition}{\bf Definition}
\newtheorem{nnlemma}{\bf Lemma}
\newtheorem{nncorollary}{\bf Corollary}
\newtheorem{nnproposition}{\bf Proposition}
\newtheorem{nexample}{\bf Example}

\newcommand{\range}{\mathrm{Ran}}

\newenvironment{theorem}
{\begin{nntheorem}\it}
{\end{nntheorem}}

\newenvironment{definition}
{\begin{nndefinition}\it}
{\end{nndefinition}}

\newenvironment{nnexample}
{\begin{nexample}\rm}{\end{nexample}}

\newtheorem{nnremark}{\bf Remark}

\def\RR{\mathbb{R}}

\def\epsilon{\varepsilon}

\newcommand{\R}{\ensuremath{\mathbb{R}}}

\renewcommand{\leq}{\leqslant}
\renewcommand{\geq}{\geqslant}

\newcommand{\A}{\mathcal{A}}
\newcommand{\Id}{I}

\title{Forwarding design for stabilization of a
coupled transport equation-ODE with a cone-bounded input nonlinearity}

\author{Swann Marx$^{1}$,  Lucas Brivadis$^{2}$ and Daniele Astolfi$^{2}$
\thanks{$^{1}$ Swann Marx is with LS2N, \'Ecole Centrale de Nantes $\&$ CNRS UMR 6004, F-44000 Nantes, France.
{\tt\small swann.marx@ls2n.fr}.}
\thanks{$^{2}$ Lucas Brivadis and Danie le Astolfi 
are with Univ Lyon, Universit\'e Claude Bernard Lyon 1, CNRS, LAGEPP UMR 5007, 43 boulevard du 11 novembre 1918, F-69100, Villeurbanne, France.
       {\tt\small lucas.brivadis@gmail.com},  {\tt\small daniele.astolfi@univ-lyon1.fr}.}%

\thanks{This research was partially supported by the French Grant ANR ODISSE (ANR-19-CE48-0004-01) and was also conducted in the framework of the regional programme "Atlanstic 2020, Research, Education and Innovation in Pays de la Loire“, supported by the French Region Pays de la Loire and the European Regional Development Fund.}
}

\begin{document}

\maketitle
\begin{abstract}
We propose a new design technique
for the stabilization of coupled ODE-PDE systems
in feedforward form. In particular, 
we address the stabilization problem 
of a one-dimensional 
transport equation
driven by  a scalar 
ODE which is controlled via a 
cone-bounded nonlinearity.
The unforced transport equation is conservative but not asymptotically stable.
The proposed technique is 
inspired 
by the forwarding approach early introduced in the 90's.
Well-posedness and asymptotic stability of the closed-loop system
are discussed.
\end{abstract}



\section{Introduction}

The study of cascade systems, also 
denoted as systems in \emph{feedforward}  form,
dates back in the 90's 
due to a large range of control applications.
A typical example of systems in this form is 
the cart-pendulum system \cite{mazenc1996adding}, 
but this structure exhibits also when adding an
integral action
for stabilization purposes \cite{astolfi2017integral,terrand2019adding}. 
In the case of ordinary differential equations 
(ODE for short), two different, although similar, techniques 
have been mainly developed:
the so-called forwarding design  \cite{mazenc1996adding, kaliora2004nonlinear,benachour2013forwarding},
and the 
nested-saturation control \cite{teel1996nonlinear}.
Both techniques have been proved to be effective 
for the stabilization of cascade
systems
 in which the control input acts, possibly via a nonlinearity or
 a saturation function, in the first subsystem-dynamics, which is supposed to be stable, and  the  output  
 of the first subsystem 
 acts on the second one, which is only marginally stable.
The goal of this work is therefore to 
propose a preliminary result 
for the stabilization of 
systems in feedforward form 
in which the second subsystem $w$
is described in particular by a ``marginally stable'' but not asymptotically stable
partial differential equation (PDE for short).
In particular, we focus here on a 
transport equation coupled
at the boundary with a scalar
ODE.
The control acts on the ODE through a cone-bounded nonlinear
function, as in the case of an actuator with dynamics. 


At the best of our knowledge, there exist few generic design techniques for the stabilization of PDEs, even less for coupled PDE-ODE systems. Let us however mention the \emph{backstepping} technique, introduced first in \cite{russell1978controllability} for first order hyperbolic PDEs, and then extended by Miroslav Krstic and his co-authors, leading to the  textbook \cite{krstic_smyshlyaev_backstepping}. This technique has been applied on some coupled PDE-ODE systems (see e.g., \cite{auriol2018delay} or \cite{tang2011state}). It is also worthy mentioning the construction of Lyapunov functionals 
with Legendre polynomials for coupled PDE-ODE systems, which leads to a hierarchy of Linear Matrix Inequalities (LMIs), and that has been applied on many systems such as the transport-ODE system \cite{safi2017tractable}, the wave-ODE system \cite{barreau2018lyapunov} or the heat-ODE system \cite{baudouin2019stability}.  Note however that in most of the mentioned examples, the control acts directly at the boundary of the PDE. In other words, a dynamic of the actuator is in general neglected in the PDE context.
A recent work  addressed the case of a coupled ODE-PDE-ODE system \cite{bribiesca}. Therein,  the feedback law is designed 
by merging  backstepping method and frequential approaches, but the the control is supposed to be linear with respect to the state. 
We finally want to emphasize on the fact that the forwarding technique has already been extended in the case of systems of first order hyperbolic PDEs, as illustrated in \cite{terrand2019adding}, but with a different viewpoint. To be more precise, the forwarding strategy has been used in order to build in a systematic way PI controllers in the one-dimensional context.

Taking into account saturations in the context of stabilization of infinite-dimensional systems is a topic that has been initiated in \cite{slemrod1989mcss} for abstract systems, where an infinite-dimensional version of the LaSalle's invariance principle is applied in order to prove the asymptotic stability of the closed-loop systems. Since then, some extensions of this result have been given, such as \cite{seidman2001note}, \cite{lasiecka2002saturation}, or more recently in \cite{marx2018stability}, which focuses on a larger class of saturations. This topic has been also studied for specific PDEs, such as the wave equation (see e.g., \cite{prieur2016wavecone}, \cite{chitour2019p}) or the 
Korteweg-de Vries equation \cite{mcpa2017siam}. To the best of our knowledge, very few is known about coupled PDE-ODE systems with saturated controllers, but let us mention anyway some recent results, such as \cite{daafouz2014nonlinear, kang2017boundary}.

In this work we address the problem of a marginally stable 
transport equation,
for which the open loop is not asymptotically stable,
coupled with an 
ODE and we propose a state-feedback design which is inspired by the forwarding technique developed for nonlinear ODEs and explained in details in \cite{mazenc1996adding,kaliora2004nonlinear,astolfi2017integral}.
In contrast with standard backstepping approaches, the gain of the 
proposed infinite-dimensional feedback law is explicit and can be
easily numerically obtained. We prove that the origin of the closed-loop system is globally asymptotically stable by applying an infinite-dimensional version of the LaSalle's invariance principle. We want to emphasize on the fact that, because of this strategy, we cannot provide any convergence rates of our trajectories, which is in contrast with the backstepping technique. Indeed, in general, an arbitrary large decay rate can be imposed in the backstepping context. However, to the best of our knowledge, there is no \emph{global} result in the saturated case for this technique, but rather local result (see e.g., \cite{kang2017boundary}).


\medskip
\noindent
\textbf{Notation:}
Set $\RR_+ = [0,\infty)$. For a function $w:\: (t,x)\in\mathbb{R}_+\times [0,1]\mapsto w(t,x)\in\mathbb{R}$, the notation $w_t$ (resp. $w_x$) denotes the partial derivative of $w$ with respect to the variable $t$ (resp. with respect to the variable $x$). We keep the notation for the weak and the strong definition of partial derivatives. When a function $w$ depends only on the variable of the time $t$ (resp. space $x$), we denote its derivative $\dot{w}$ (resp. $w^\prime$). 
Set $\Omega = \R_+\times(0, 1)$.
Let $L^2(0, 1)$ be the Hilbert space of real-valued square-integrable functions over the interval $(0,1)$. We denote $\langle f, g\rangle_{L^2} = \int_0^1fg$ the inner product between $f,\, g\in L^2(0,1)$ and $\|f\|_{L^2}$ the induced norm.
Let $H^1(0, 1)\subset L^2(0,1)$ be the Hilbert space of real-valued absolutely continuous functions over $[0, 1]$ with square-integrable derivative. We denote $\langle f, g\rangle_{H^1} = \langle f, g\rangle_{L^2}+\langle f', g'\rangle_{L^2}$ the inner product between $f,\, g\in H^1(0,1)$ and $\|f\|_{H^1}$ the induced norm.

\section{Main Results}
\label{sec_main}

\subsection{Problem Statement}

In this paper, we are concerned with the stabilization problem of
systems in feedforward form  of the form
\begin{equation}
\label{open-loop}
\left\{
\begin{array}{ll}
\dot{z} (t) = - a z(t) + \sigma(u(t))& \forall t\in\R_+\\
w_t(t,x) + \lambda w_x(t,x) = 0& \forall (t,x)\in\Omega\\
w(t,0) = w(t,1) + \gamma z(t)& \forall t\in\R_+\\
z(0)=z_0,\ w(0,x) = w_0(x)& \forall x\in(0,1),
\end{array}
\right.
\end{equation}
where $z(t)\in\mathbb{R}$ and $w(t,x)\in\mathbb{R}$ are the \emph{states} of the system, $u(t)\in\mathbb{R}$ is the control,
 $a>0$, $\gamma\neq0$, and 
 $\lambda$ is  the \emph{velocity} of the transport equation. The parameters
 $a,\gamma,\lambda$ are supposed to be known and the function $\sigma:\RR\to\RR$ is a \emph{cone-bounded function} satisfying 
the definition below.
System \eqref{open-loop} can represent, for instance, a  transport equation
with an actuator admitting a dynamics, represented by the state $z$.
In contrast with most of the works in the literature, 
we consider here the case of a transport equation which is
``marginally stable''. In fact, due to the boundary condition $w(t,1)=w(t,0)$,
trajectories of the ``autonomous'' $w$-dynamics, i.e. with $z=0$, are not diverging nor converging to zero. 


\begin{definition}[Cone-bounded function]
\label{def-sat}
A continuous function $\sigma:\: \mathbb{R}\rightarrow \mathbb{R}$ is said to be \emph{cone-bounded} if the following holds.
\begin{enumerate}
\item For all $s\in\R$, $\sigma(s)=0$ if and only if $s=0$.
\item It is nondecreasing, i.e., for all $(s_1,s_2)\in\mathbb{R}^2$,
$$(\sigma(s_1)-\sigma(s_2))(s_1-s_2)\geq 0.$$  
\item It is globally Lipschitz, i.e. there exists $m>0$ such that for all $(s_1, s_2)\in\mathbb{R}^2$,
\begin{equation}
|\sigma(s_1)-\sigma(s_2)|\leq m|s_1-s_2|.
\end{equation}
\end{enumerate}
\end{definition}

Some examples of satisfying the definition of 
cone-bounded functions are the following:
linear maps of the form $\sigma_1:s\mapsto \rho s$ for some $\rho\neq0$;
saturation functions such as
$$
\sigma_2:s\mapsto\begin{cases}
\rho s_1&\text{if } s\leq s_1\\
\rho s &\text{if } s_1\leq s\leq s_2\\
\rho s_2&\text{if } s_2\leq s
\end{cases}
$$
for some $\rho\neq0$, $s_1<s_2$ or $\sigma_3:s\mapsto\theta\,\mathrm{arctan}(\rho s)$
for some $\rho,\,\theta\neq0$.


\subsection{Forwarding Design and Stability Analysis}

The aim of this paper is
to propose an extension of the \emph{forwarding} technique, developed first for finite dimensional systems (see e.g., \cite{mazenc1996adding} for more details) 
to stabilize the origin of system
\eqref{open-loop}, which is infinite-dimensional because of the transport equation. Note that the extension is not trivial since it requires to carefully study the regularity of the solutions.

Without loss of generality, we suppose that the $z$-dynamics is already stable when $u=0$, since  
this condition is necessary for the stabilizability of the overall cascade system. 
The $z$-dynamics is therefore considered as the fast-dynamics while the 
$w$ represents the slow-dynamics. As a consequence, the main 
strategy consists in designing a feedback law able to 
stabilize the $w$-dynamics on an 
invariant manifold that depends on $z$, which is chosen, in light of the linear properties of the dynamics of $z$ and $w$ as $Mz$ with $M:[0,1]\to\RR$
a function to be defined. 
According to such strategy, we select
the feedback law as
\begin{align}
u(t) &= \mu \int_0^1 (w(t,x) - M(x) z(t))M(x) dx\nonumber\\
&= \mu \big\langle w(t) - Mz(t)), M \big\rangle_{L^2} \label{feedback}
\end{align}
with $\mu$ a positive constant.
Note that if $w(\cdot,x)-M(x)z$ converges to zero, then $u$ converges to zero, and since $z$ converges to zero when $u=0$, one can hope that both $w$ and $z$ converges to zero if $M$ is defined in an appropriate way.
To this end, we
define $M:[0,1]\to \RR$ as the solution of the two-point boundary value problem
\begin{equation}
\label{ODE-M}
\left\{
\begin{aligned}
&a M(x) = \lambda M^\prime(x)\\
& M(0) = M(1) + \gamma.
\end{aligned}
\right. 
\end{equation}
It is straightforward to see that the solution to the 
latter ODE is given by the formula
\begin{equation}
\label{expression-M}
M(x) = \frac{\gamma \exp\left(\frac{a}{\lambda} x\right)}{1-\exp\left(\frac{a}{\lambda}\right)},\qquad \forall x\in [0,1].
\end{equation}
System \eqref{open-loop} in closed-loop with the feedback-law \eqref{feedback} 
reads, for all for $(t, x)\in\Omega$,
\begin{equation}
\label{closed-loop}
\left\{
\begin{array}{ll}
\dot{z} (t) = -a z(t)
+ \sigma\left( \mu \langle w(t) - M z(t), M \rangle_{L^2}\right) 
\\
w_t(t,x) + \lambda w_x(t,x) = 0
\\
w(t,0) = w(t,1) + \gamma z(t)
\\
z(0)=z_0,\ w(0,x) = w_0(x)
\end{array}
\right.
\end{equation}
The following result ensures that there exists a unique solution to the  Cauchy problem
for system \eqref{closed-loop}.

\begin{theorem}[Well-posedness and Lyapunov Stability]
\label{thm-wp}
Let $(z_0,w_0)\in \R\times L^2(0, 1)$. Then \eqref{closed-loop} has a unique solution $(z,w)\in C^0((0,\infty);\R\times L^2(0, 1))$ such that $(z(0), w(0)) = (z_0,w_0)$. Moreover, there exists $\alpha>0$ such that for all $t\in\R_+$,
\begin{equation}
\label{bounded-H}
|z(t)| + \Vert w(t,\cdot)\Vert_{L^2} \leq \alpha(|z_0|+\Vert w_0\Vert_{L^2}).
\end{equation}
Furthermore, if $(z_0,w_0)\in \R\times H^1(0,1)$ satisfies the compatibility condition
\begin{equation}\label{compatibility}
w_0(0) = w_0(1) + \gamma z_0,
\end{equation}
then \eqref{closed-loop} has a unique solution $(z,w)\in C^0((0,\infty);\R\times H^1(0,1)) \cap C^1((0,\infty);\R\times L^2(0,1))$
such that $(z(0), w(0)) = (z_0,w_0)$.
and there exists $\alpha>0$ such that for all $t\in\R_+$,
\begin{equation}
\label{bounded-D(A)}
\begin{aligned}
|z(t)| + \Vert w(t,\cdot)\Vert_{H^1} \leq \alpha(|z_0|+\Vert w_0\Vert_{H^1}).
\end{aligned}
\end{equation}
\end{theorem}

Previous theorem is crucial, not only to ensure the well-posedness of the closed-loop system \eqref{closed-loop}, but also because  inequalities \eqref{bounded-H} and \eqref{bounded-D(A)} imply the Lyapunov stability at the origin of the system. 
As a consequence, it remains to show the attractivity of the origin to ensure the global asymptotic stability, which is the second main result of the paper.

\begin{theorem}
\label{thm-as}
Let $(w_0,z_0)\in L^2(0,1)\times\mathbb{R}$. Then the origin of \eqref{closed-loop} is globally asymptotically stable in the $L^2(0,1)\times \mathbb{R}$-topology. 
\end{theorem}

\section{Proofs of Theorems}
\label{sec_proof}

\subsection{Proof of Theorem \ref{thm-wp}}
\label{sec_proof1}

The proof of Theorem \ref{thm-wp} relies on the so-called semigroup theory (see, e.g., \cite{tucsnak2009observation} or \cite{pazy1983semigroups} for an introduction to this theory in the linear case, and \cite{miyadera1992nl_sg} in the nonlinear case). 
Let $X=\R\times L^2(0,1)$ be the state space of \eqref{closed-loop}. It is a Hilbert space as the Cartesian product of two Hilbert spaces. For all $((z_1, w_1), (z_2, w_2))\in X^2$, set
\begin{align}
\big\langle (z_1, w_1),& (z_2, w_2) \big\rangle_X\label{psX}\\
&=
a z_1z_2 + \mu m \textbf{}\langle w_1 - Mz_1, w_2-Mz_2 \big\rangle_{L^2}.
\nonumber
\end{align}
Then, $\langle\cdot,\cdot\rangle_X$ defines an inner product on $X$ that induces a norm $\|\cdot\|_X$ that is equivalent to the norm induced by the Cartesian product. Indeed, for all $(z, w)\in X$,
\begin{align}
&\|(z, w)\|^2_X\\
&= a |z|^2 + \mu m\left( \|w\|^2_{L^2}
- 2z\langle M,w \rangle_{L^2}
+ |z|^2\|M\|_{L^2}^2\right).\nonumber
\end{align}
Hence, applying Cauchy-Schwarz and Young's inequalities, one can find 
positive constants $\underline\alpha,\ \bar{\alpha}>0$ such that 
\begin{align*}
\underline\alpha\left(|z|^2 + \|w\|^2_{L^2}\right)
\leq \|(z, w)\|^2_X
\leq \bar\alpha\left(|z|^2 + \|w\|^2_{L^2}\right).
\end{align*}
Now, let us define the following nonlinear unbounded operator
\begin{equation}
\label{operator-A}
\begin{aligned}
\mathcal{A}: D(\mathcal{A})&\rightarrow X\\
(z, w) &\mapsto
(-a z + \sigma\left(\mu\langle w-M z, M \rangle_{L^2}\right) , -\lambda w^\prime)
\end{aligned}
\end{equation} 
where $D(\mathcal{A})$, called the \emph{domain of $\mathcal{A}$}, is given by
$$
D(\mathcal{A})=\lbrace (z,w)\in
\R\times H^1(0, 1)
\mid w(0)=w(1) + \gamma z\rbrace,
$$
equipped with the graph norm: $\Vert \cdot\Vert_{D(\mathcal{A})}=\Vert \cdot\Vert_X + \Vert \mathcal{A} (\cdot)\Vert_X$.
Then $\mathcal{A}$ is densely defined on $X$.
Thanks to this operator, and denoting by $\zeta = (z, w)$ the state of the system,
one can re-write \eqref{closed-loop} as an abstract Cauchy problem
\begin{equation}
\dot{\zeta} = \mathcal{A}(\zeta), 
\qquad \zeta(0)=\zeta_0\in X.
\end{equation}
In order to prove Theorem~\ref{thm-wp}, we show that $\mathcal{A}$ is the generator of a strongly continuous contraction semigroup over $X$, denoted by $(\mathbb{T}(t))_{t\geq 0}$.
If one proves that the operator $\mathcal{A}$ defined in \eqref{operator-A} is a $m$-dissipative operator on $(X, \|\cdot\|_X)$, 
according to the definition below,
one can apply the result provided by \cite[Corollary 3.7, Theorem 4.20]{miyadera1992nl_sg},
and conclude that the statements of Theorem \ref{thm-wp} hold true.
\begin{definition}[$m$-dissipative operators]\label{def_dissipative}
An operator $\mathcal{A}:\: D(\mathcal{A})\subset X\rightarrow X$ is said to be  $m$-dissipative if 
\begin{enumerate}[left= 0pt]
\item The operator $\mathcal{A}$ is \emph{dissipative}, i.e.
\begin{equation}\!\!\!
\big\langle \mathcal{A}(\zeta_1)-\A(\zeta_2),\zeta_1-\zeta_2
\big\rangle_X\leq 0, \;\,
\forall (\zeta_1,\zeta_2)\in D(\mathcal{A})^2.
\end{equation}
\item The operator $\mathcal{A}$ is maximal, i.e. there exists $\lambda_0>0$ such that
\begin{equation}
\range(\lambda_0\Id_X - \mathcal{A})=X,
\end{equation}
where $I_X$ is the identity operator over the Hilbert space $X$, and $\range$ is the range operator.
\end{enumerate}
\end{definition} 

As a consequence, the rest of the proof consists in 
showing that the operator $\mathcal{A}$ 
defined in \eqref{operator-A}
is $m$-dissipative according to Definition~\ref{def_dissipative}.

\textbf{Step 1: $\mathcal{A}$ is dissipative.}
Let $\zeta_1=(z_1, w_1),\, \zeta_2=(z_2, w_2)\in X$.
Set $\tilde{z} = z_1-z_2$, $\tilde{w} = w_1-w_2$ and $u_i = \mu\langle w_i-Mz_i, M\rangle_{L^2}$ for $i\in\{1,2\}$.
Using the definition of the domain of $\A$ and the boundary conditions of \eqref{ODE-M}, we get
$$
\big\langle \tilde{w}^\prime - M^\prime\tilde{z}, \tilde{w} - M \tilde{z}\big\rangle_{L^2}
= 0.
$$
Then, by using the definition 
of $M$ in \eqref{ODE-M} and equation \eqref{psX}, 
we can derive
\begin{align}
\big\langle \A &(\zeta_1)-\A(\zeta_2), \zeta_1-\zeta_2\big\rangle_X
\nonumber\\
&\qquad =\big\langle
(-a\tilde{z}+\sigma(u_1)-\sigma(u_2), -\lambda \tilde{w}^\prime),
(\tilde{z}, \tilde{w})\big\rangle_X
\nonumber\\
&\qquad = -a^2 \tilde{z}^2 + a\tilde{z}(\sigma(u_1)-\sigma(u_2))
\nonumber\\
&\qquad \quad- m(\sigma(u_1)-\sigma(u_2))(u_1-u_2)
\nonumber\\
&\qquad \leq -\frac{a^2}{2}\tilde{z}^2
-\frac{m}{2}(\sigma(u_1)-\sigma(u_2))(u_1-u_2) \nonumber
\\
&\qquad \leq 0\label{lyap_ineq}
\end{align}
since $\sigma$ is $m$-Lipschitz and non-increasing.

\textbf{Step 2: $\mathcal{A}$ is maximal.} 
Proving that $\A$ is maximal reduces to show that 
for all $\zeta\in X$, there exists $\tilde{\zeta}\in D(\mathcal{A})$ such that
\begin{equation}
(I_X-\mathcal{A})\tilde{\zeta} = \zeta.
\end{equation} 
Let $(z, w)\in X$.
We seek $(\tilde{z}, \tilde{w})\in D(\A)$ such that
\begin{equation}
\label{system-maximal}
\left\{
\begin{aligned}
&\tilde{z} + a \tilde{z} - \sigma(u)=z\\
& \tilde w + \lambda \tilde{w}^\prime(x) = w
\end{aligned}
\right.
\end{equation}
where $u=\mu\langle\tilde{w} - M \tilde{z}, M\rangle_{L^2}$. 
Assume that $(\tilde{z}, \tilde{w})\in D(\A)$ satisfies \eqref{system-maximal}.
Thanks to the variation of constants formula,
\begin{align}
\tilde w(x) &= \tilde w(0)\exp(-\lambda^{-1} x)\nonumber\\
&\quad+\lambda^{-1}\int_0^x \exp(-\lambda^{-1} (x-s)) w(s) ds
\label{eq_wexp}
\end{align}
for all $x\in[0, 1]$.
Using the boundary conditions, we get that
\begin{equation}\label{eq_bc}
\tilde w(0)=\frac{\gamma \tilde{z} + \lambda^{-1}\int_0^1 \exp(-\lambda^{-1} (1-s))w(s) ds}{1-\exp(-\lambda^{-1})}.
\end{equation}
Combining \eqref{eq_wexp} and \eqref{eq_bc}, we get an expression of $\tilde w$ depending only on $\tilde z$ and $w$.
For all $x\in[0, 1]$, set
\begin{align*}
    &G(x)=\lambda^{-1} \int_0^x \exp(-\lambda^{-1} (x-s)) w(s) ds\\
    &+ \frac{\lambda^{-1}\exp(-\lambda^{-1} x)}{1-\exp(-\lambda^{-1})}\int_0^1 \exp(-\lambda^{-1}(1-s)) w(s) ds 
\end{align*}
and
\begin{align*}
C &= \frac{\mu\gamma}{1-\exp(-\lambda^{-1})}\int_0^1\exp(-\lambda^{-1} x) M(x) dx\\&\quad - \mu \int_0^1 M(x)^2 dx.
\end{align*}
Then
\begin{equation}
u = - C \tilde{z} + \langle M, G\rangle_{L^2}. 
\end{equation}
According to \eqref{expression-M}, $\gamma M(x)\leq0$ for all $x\in(0,1)$. Hence $C\leq 0$.
Consider the map $f: \R\ni\tilde z\mapsto\tilde z + a \tilde z - \sigma(-C\tilde z + \langle M, G\rangle_{L^2})$.
Since $C \leq 0$ and $\sigma$ is continuous and non-decreasing, 
$f$ is also continuous and we have
$\lim_{\tilde{z}\rightarrow + \infty}f(\tilde{z})= + \infty$ and $\lim_{\tilde{z} \rightarrow - \infty } f(\tilde z)= -\infty$.
Thus, for all $(z, w)\in X$, according to the intermediate value theorem, there exist $\tilde z$ such that $f(\tilde z) = z$.
Then, the pair $(\tilde z, \tilde w)$ where $\tilde w$ satisfies  \eqref{eq_wexp} and \eqref{eq_bc} is solution of \eqref{system-maximal}.
Thus, the operator $\A$ is maximal.\qed

\subsection{Proof of Theorem \ref{thm-as}}
\label{sec_proof2}
We prove the statement of the theorem for initial conditions $(z_0, w_0)$ in $\R\times H^1(0, 1)$ satisfying the compatibility condition \eqref{compatibility}.
The result easily follows for all initial conditions in $\R\times L^2(0, 1)$ by a standard density argument (see, e.g., \cite[Lemma 1]{map2017mcss})).

The proof of Theorem \ref{thm-as} relies on the use of the LaSalle's invariance principle. One has to be careful in the infinite-dimensional case, because it is necessary to check whether the positive orbit is precompact (see, e.g., \cite{map2017mcss} or \cite{slemrod1989mcss} for more details). This is the case for the positive orbit associated to the dynamical system \eqref{closed-loop}, since the canonical embedding from $D(\mathcal{A})$ equipped with the graph norm into $X$ is compact. We can now prove Theorem \ref{thm-as}.

Let $(z_0,w_0)\in D(\mathcal{A})$ satisfying the compatibility condition \eqref{compatibility}. According to Theorem~\ref{thm-wp},  system \eqref{closed-loop} has a unique solution $(z,w)\in C^0((0,\infty);\R\times H^1(0,1)) \cap C^1((0,\infty);\R\times L^2(0,1))$ with initial condition $(z_0, w_0)$.
Let $u = \mu\langle w-Mz, M\rangle_{L^2}$.
Using the dissipativity of $\A$, we get
\begin{align}
\frac{1}{2}\frac{d}{dt}\|(z, w)\|_X^2
&= \big\langle A(z, w), (z, w)\big\rangle_X
\nonumber
\\
&\leq -a^2 z^2
-m\sigma(u)u.
\tag{by \eqref{lyap_ineq}}
\end{align}
Then, for all $t\in\R_+$,
\begin{align*}
    \|(z(t), w(t))\|_X^2 &- \|(z_0, w_0)\|_X^2\\
    &\leq
    - \int_0^t a^2 z(s)^2 + m \sigma(u(s))u(s) ds.
\end{align*}
Then,
\begin{equation*}
\begin{aligned}
\int_0^{+\infty} z(s)^2 ds <+\infty,\quad
\int_0^{+\infty} \sigma(u(s))u(s) ds <+\infty.
\end{aligned}
\end{equation*}
The Lyapunov stability given by \eqref{bounded-D(A)} together with the fact that $\sigma$ is globally Lipschitz imply that $z$ and $s\mapsto\sigma(u(s))u(s)$ are also globally Lipschitz. Hence, applying Barbalat's lemma, we have
\begin{equation}
\lim_{t\rightarrow +\infty} z(t) =0,\quad \lim_{t\rightarrow + \infty} \sigma(u(t))u(t) = 0
\end{equation}
Since $\sigma$ is a cone-bounded function, $u(t)\to0$ as $t\to+\infty$ (Otherwise, take a subsequence of $u$ and use that $\sigma$ vanishes only at $0$). 
Hence,
\begin{equation}
\lim_{t\rightarrow +\infty} z(t) =0,\quad \lim_{t\rightarrow + \infty} \big\langle w(t), M \big\rangle_{L^2} = 0.
\end{equation}
Recall that the $\omega$-limit set of the initial condition $(z_0, w_0)$, denoted by $\omega(z_0, w_0)$, is the set of all $(z^\star, w^\star)\in D(\A)$ such that there exists an increasing sequence of time $(t_n)_{n\geq0}$ such that $z(t_n)\to z^\star$ and $w(t_n)\to w^\star$ in $L^2(0, 1)$ as $n$ goes to infinity.
We are going to prove that $\omega(z_0, w_0)$ reduces to $(0, 0)$, so that $(z, w)$ converges to the origin in the $\R\times L^2(0, 1)$-topology since it is bounded according to \eqref{bounded-D(A)}.
Since $(z, w)$ is uniformly bounded in time according to \eqref{bounded-D(A)} and the canonical inclusion $D(\A)\hookrightarrow X$ is compact according to the Sobolev inclusion theorems, the positive orbit $\{(z(t), w(t)),\ t\in\R_+\}$ is precompact in $X$.
Therefore, according to the LaSalle's invariance principle for infinite-dimensional systems (see \cite[Theorem 3.1]{slemrod1989mcss}), $\omega(z_0, w_0)$ is a non-empty compact subset of $X$ that is invariant to the flow of \eqref{closed-loop}.
Hence, for any initial condition $(z^\star_0, w^\star_0)\in\omega(z_0, w_0)$,
the corresponding solution $(z^\star, w^\star)\in C^0((0,\infty);\R\times H^1(0,1)) \cap C^1((0,\infty);\R\times L^2(0,1))$ of \eqref{closed-loop} satisfy
\begin{equation}
\label{omega-equation}
\left\{
\begin{array}{ll}
w^\star_t + \lambda w^\star_x = 0&\\
w^\star(t,0) = w^\star(t,1),& \forall t\in\mathbb{R}_+\\
\big\langle w^\star(t), M\big\rangle_{L^2} = 0,& \forall t\in\mathbb{R}_+\\
w^\star(0) = w^\star_0, \\
z^\star = 0 &
\end{array}
\right.
\end{equation}
Using \eqref{ODE-M} and performing an integration by parts, we get
\begin{align*}
   \big \langle w^\star(t), M\big\rangle_{L^2}
    &= \tfrac{\lambda}{a}\big\langle w^\star(t), M'\big\rangle_{L^2}\\
    &= - \tfrac{\lambda}{a}\big\langle w_x^\star(t), M\big\rangle_{L^2}
    - \tfrac{\gamma\lambda}{a}w^\star (t, 0)\\
    &= - \tfrac{1}{a}\big\langle w_t^\star(t), M\big\rangle_{L^2}
    - \tfrac{\gamma\lambda}{a}w^\star (t, 0)\\
    & =  - \tfrac{\gamma\lambda}{a}w^\star(t, 0)
\end{align*}
since $\langle w^\star, M\rangle_{L^2} = 0$.
Then, $w^\star$ satisfies
\begin{equation}
\label{system-omega2}
\left\{
\begin{array}{ll}
w^\star_t + \lambda w^\star_x = 0&\\
w^\star(t,0) = w^\star(t,1) = 0,& \forall t\in\mathbb{R}_+\\
\end{array}
\right.
\end{equation}
Let us define $E_1(w^\star) = \int_0^1 w^\star(t,x)^2 dx$ and $E_2(w^\star)=\int_0^1 e^{-x}w^\star(t,x)^2 dx$, which are candidate Lyapunov functions equivalent to the usual norm in $L^2(0,1)$.
Then, for all $t\in\R_+$,
$E_1(w^\star(t)) = E_1(w_0^\star)$, and
$E_2(w^\star(t)) = e^{-\lambda t} E_2(w_0^\star)$,
which means that the energy is preserved and also converges to $0$. Such a case occurs if and only if $w^\star=0$ in $L^2(0,1)$.
Therefore, $\omega(z_0, w_0) = \{(0,0)\}$, which means
that $(z, w)$ converge to $(0, 0)$ in the $\R\times L^2(0, 1)$-topology.
Then by a standard density argument (see, e.g., \cite[Lemma 1]{map2017mcss}) the result also holds for initial conditions $(z_0, w_0)$ in $\R\times L^2(0, 1)$, which concludes the proof of Theorem~\ref{thm-as}.\qed





\section{Numerical simulations}

To illustrate the results, let us fix the parameters and initial condition of \eqref{closed-loop} as 
    $\lambda = 1$, 
    $a = 1$,
    $\gamma = 1$,
    $\sigma = \arctan$, $z_0 = 1$,
    $w_0:x\mapsto \sin(2\pi x)-x$.

\begin{figure}
    \centering
    \includegraphics[width=.42\textwidth]{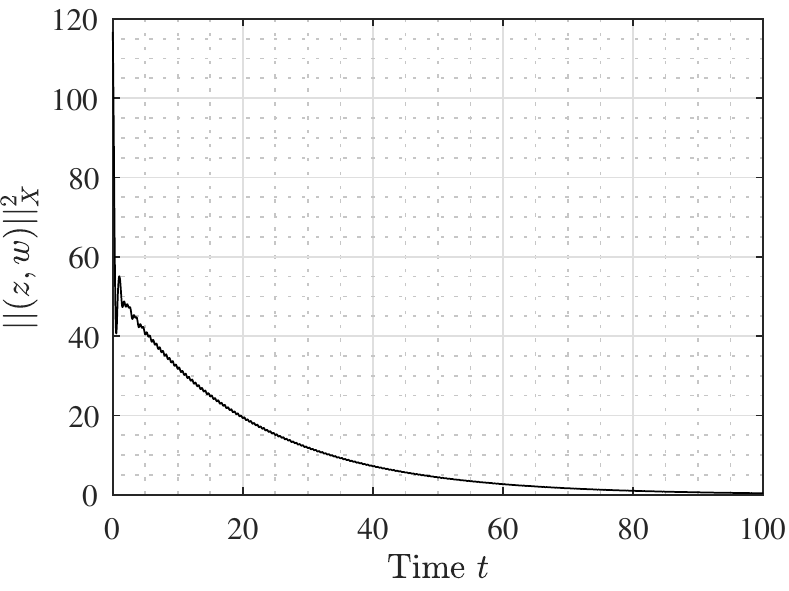}
    \caption{Evolution of $\|(z, w)\|^2_X$ on the time interval $[0, 100]$.}
    \label{fig:X}
\end{figure}

\begin{figure}
    \centering
    \begin{align*}
    \includegraphics{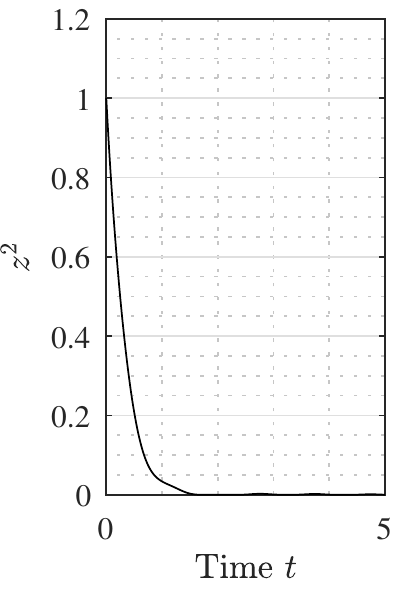}
    &
    \includegraphics{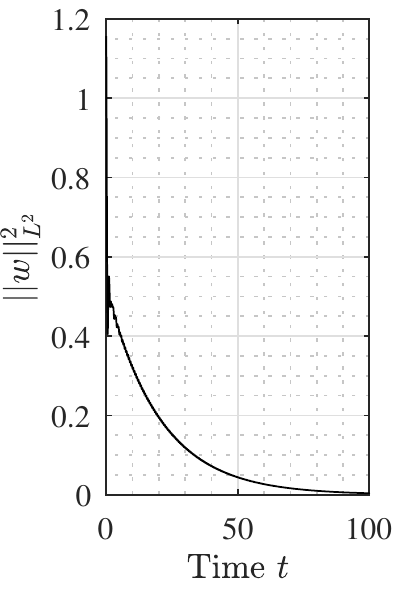}
    \end{align*}\vspace{-2em}
    \caption{Evolution of $z^2$ and $\|w\|^2_{L^2}$ on the time interval $[0, 100]$.}
    \label{fig:zw}
\end{figure}

Using the method of characteristics (see, e.g., \cite[Section 2.1]{Evans}), one can compute the exact solution of \eqref{closed-loop} at any time, and obtain the numerical results drawn on Figures~\ref{fig:X} and~\ref{fig:zw}.
As stated by Theorem~\ref{thm-as}, $\|(z, w)\|^2_X$ seems to converge to zero as time goes to infinity in Figure~\ref{fig:X}.
Moreover, Figure~\ref{fig:zw} shows that the convergence of $z$ to zero seems to be faster than the convergence of $w$ to zero.
This phenomenon is typical of forwarding design techniques when applied to cascade systems in the finite-dimensional context (see, e.g., 
\cite{mazenc1996adding,teel1992globalsaturation}).


\section{Conclusion}
\label{sec_conc}

In this paper, we have extended the forwarding technique to a coupled PDE-ODE system in a feedforward form. Our strategy relied on the introduction of a suitable Lyapunov functional and the application of the infinite-dimensional LaSalle's invariance principle.

These results pave the way to many other extensions. 
The first natural one is the application of our strategy on some systems of linear transport equations coupled with some systems of ODEs, possibly in an output feedback context. This would lead to some algebraic conditions on the matrices involved in the closed-loop, mixing Kalman rank condition together with conditions such as the ones provided in \cite{bastin2016stability}. 
An estimation of the decay rates should be studied so that to 
have a better insight in  how the parameters of the controller can improve or not the performances of the closed-loop system.  
Finally, we believe that this preliminary work may open 
the route to different approaches in the context of 
periodic regulation \cite{astolfi2019francis}, 
 repetitive control \cite{califano2018stability}
 and feedback stabilization in presence of 
 input/output delays.\\

\noindent
\textbf{Acknowledgments.} 
We thank Nathan van de Wouw for the fruitful
discussions that helped in the outcome of this work.

\bibliographystyle{plain}
\bibliography{bibsm}

\end{document}